\documentclass[aps,prl,onecolumn,superscriptaddress,showpacs,floatfix]{revtex4}

\pdfoutput=1

\usepackage{amsmath}
\usepackage{bm}
\usepackage{amsfonts}
\usepackage[pdftex]{color,graphicx}
\usepackage[pdftex]{hyperref}

\newcommand{\noEfieldtable}
{
\begin{table}[b]
    \caption{The position (height), charge transfer, and atomic
    magnetic moment of each metal atom adsorbed on large unit-cell
    graphene, and total magnetic moment of each model system. GGA 
    results are shown in the Voronoi charge scheme.} \label{Tab1}
    \begin{ruledtabular}
    \begin{tabular}{cccc}
     { }  & Cs & Ba & La  \\
     \hline
     Height (\AA)  & 3.01 & 2.65 & 2.38 \\
     Charge ($e$)  & 0.95 & 1.20 & 1.27 \\
     Atomic moment ($\mu_{\rm B}$) & 0.00 & 0.40 & 1.27 \\
     Total moment ($\mu_{\rm B}$) & 0.00 & 0.47 & 1.86  \\
   \end{tabular}
   \end{ruledtabular}
\end{table}
}

\newcommand{\Efieldtable}
{
\begin{table}[b]
    \caption{The position (height), charge transfer, and atomic
    magnetic moment of La atom adsorbed on large unit-cell graphene 
    with an external electric field. Total magnetic moment change of 
    the system also appears.} \label{Tab2}
    \begin{ruledtabular}
    \begin{tabular}{ccccc}
    field strength & height & charge & atomic moment & total moment \\
      (V/nm) & (\AA) & (e) & ($\mu_{\rm B}$) & ($\mu_{\rm B}$) \\
     \hline
     0 & 2.38 & 1.27 & 1.27 & 1.86 \\
     1 & 2.27 & 1.43 & 1.04 & 1.77 \\
     2 & 2.18 & 1.58 & 0.75 & 1.44 \\
   \end{tabular}
   \end{ruledtabular}
\end{table}
}

\newcommand{\strfig}
{
\begin{figure}[t]
   \includegraphics[width=\columnwidth]{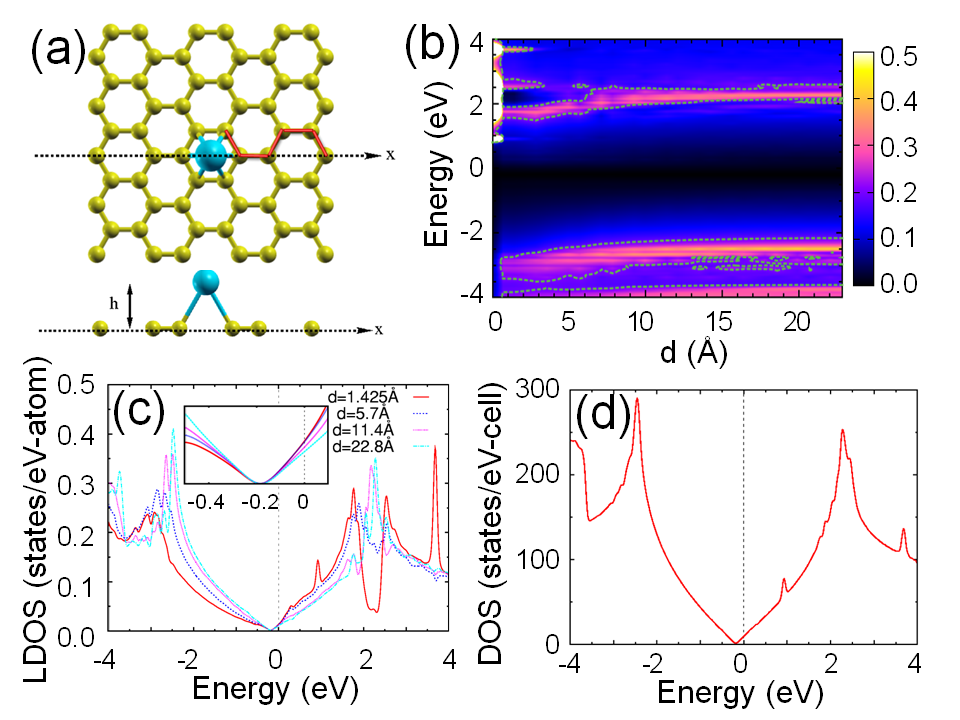}
   \caption{STS simulations for Cs-adsorbed graphene. (a) Small unit 
   cell model. (b) 2D simulated STS map obtained using the large unit 
   cell. $d$ denotes the $x$-components of the amchair line depicted 
   by the red line in (a). (c) LDOSs at distances of $d$ = 1.425, 5.7, 
   11.4, and 22.8 \AA~correspond to vertical cuts of the STS map.
   Inset shows the magnified LDOS near the Dirac point. (d) Total DOS 
   of Cs-adsorbed graphene. The Fermi level is set to zero.}
   \label{Fig1}
\end{figure}
}

\newcommand{\STSfig}
{
\begin{figure}[t]
   \centering
   \includegraphics[width=\columnwidth]{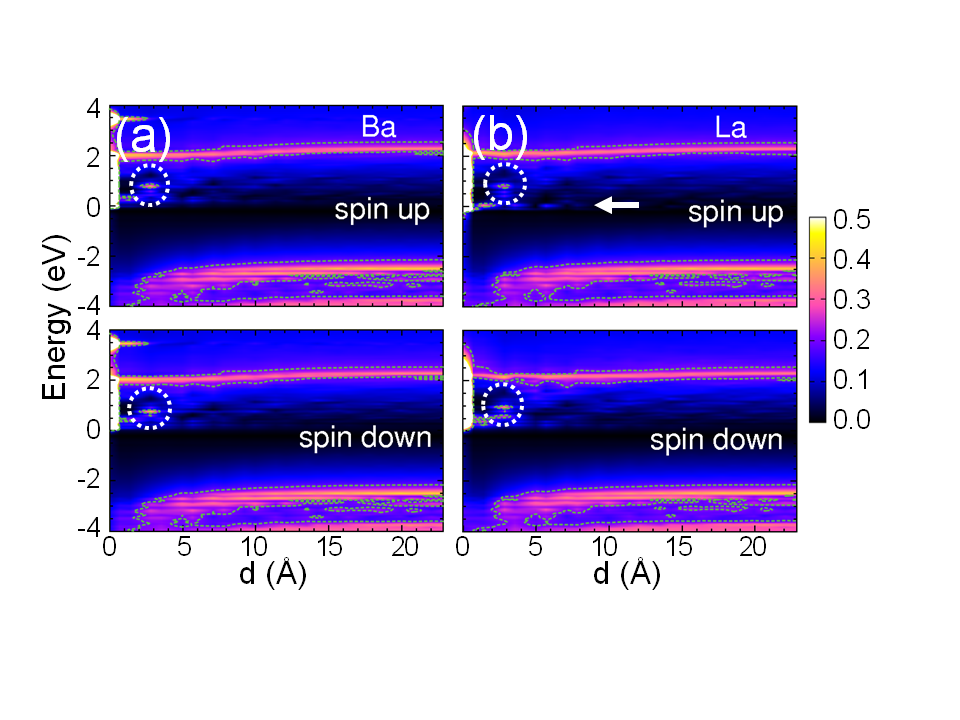}
   \caption{Simulated STS maps of (a) Ba- and (b) La-adsorbed
   graphene layers. The arrow in (b) indicates the location of the state 
   corresponding to Fig.~\ref{Fig3}(a). The dashed-lined 
   circles denote symptoms of the covalent characters.}
\label{Fig2}
\end{figure}
}

\newcommand{\Sfig}
{
\begin{figure}[t]
   \centering
   \includegraphics[width=\columnwidth]{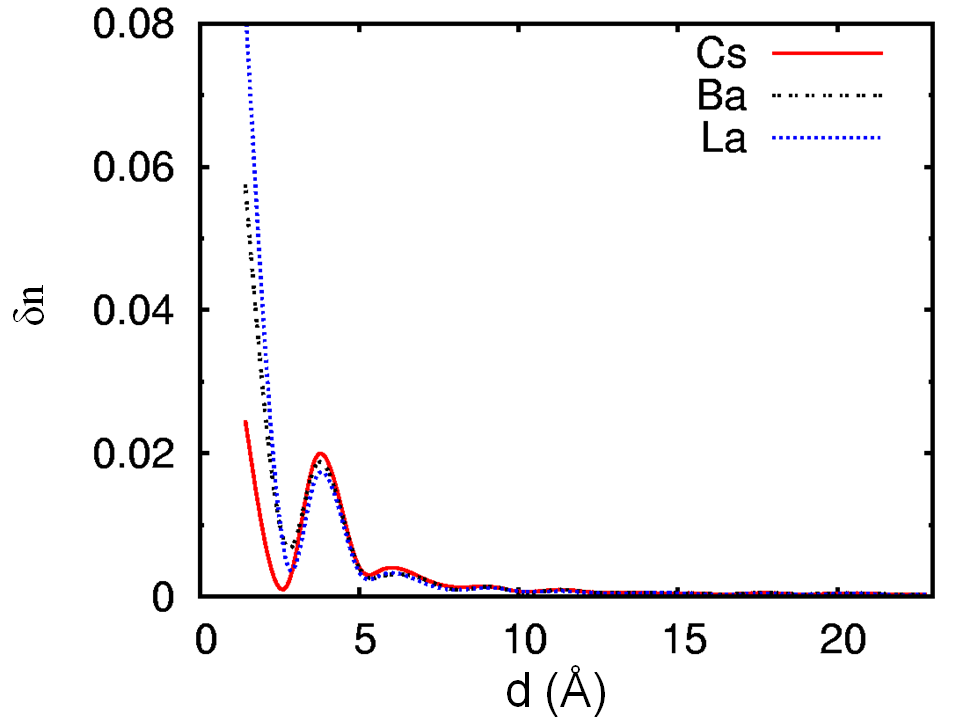}
   \caption{The Friedel oscillation patterns caused by the charge impurities
    for  Cs-, Ba- and La-adsorbed graphene layers. Within a few angstrom, the
    amplitudes of the charge density oscillaltions rapidly decrease for all
    three metal adsorbates.} 
\label{FigS1}
\end{figure}

\begin{figure}[t]
   \centering
   \includegraphics[width=\columnwidth]{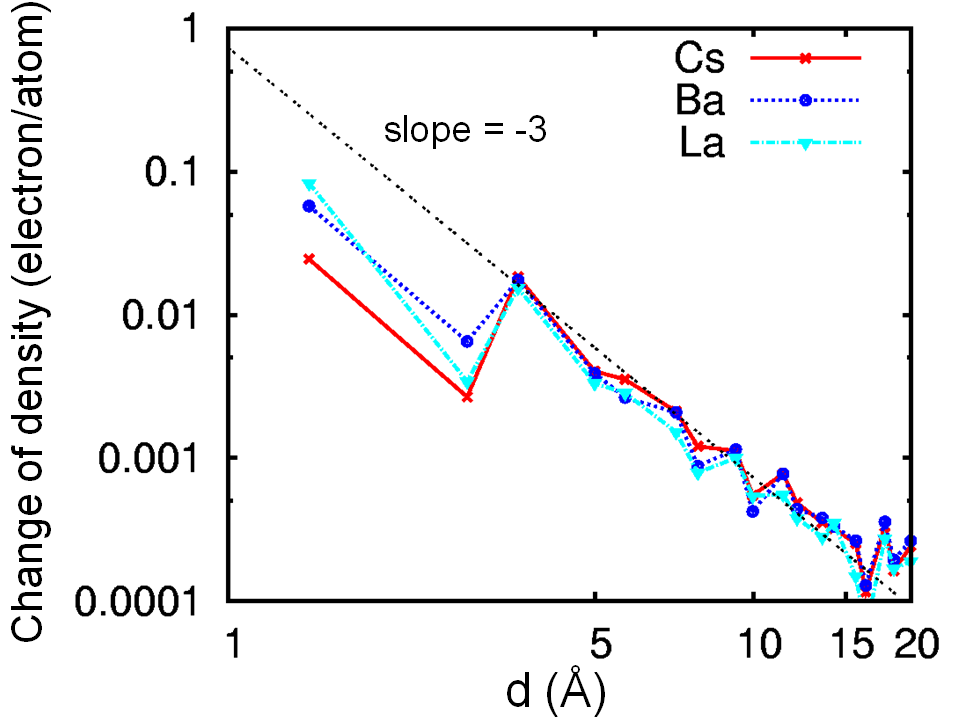}
   \caption{Changes in electronic density of the graphene layer during metal
     atom adsorption. The infinite distance limit (~0.0002 e) was already
     subtracted. Both axes are plotted on the logarithmic scale. The data were
     obtained from Voronoi charge analysis.}
\label{FigS2}
\end{figure}
}

\newcommand{\Ediafig}
{
\begin{figure}[t]
   \centering
   \includegraphics[width=\columnwidth]{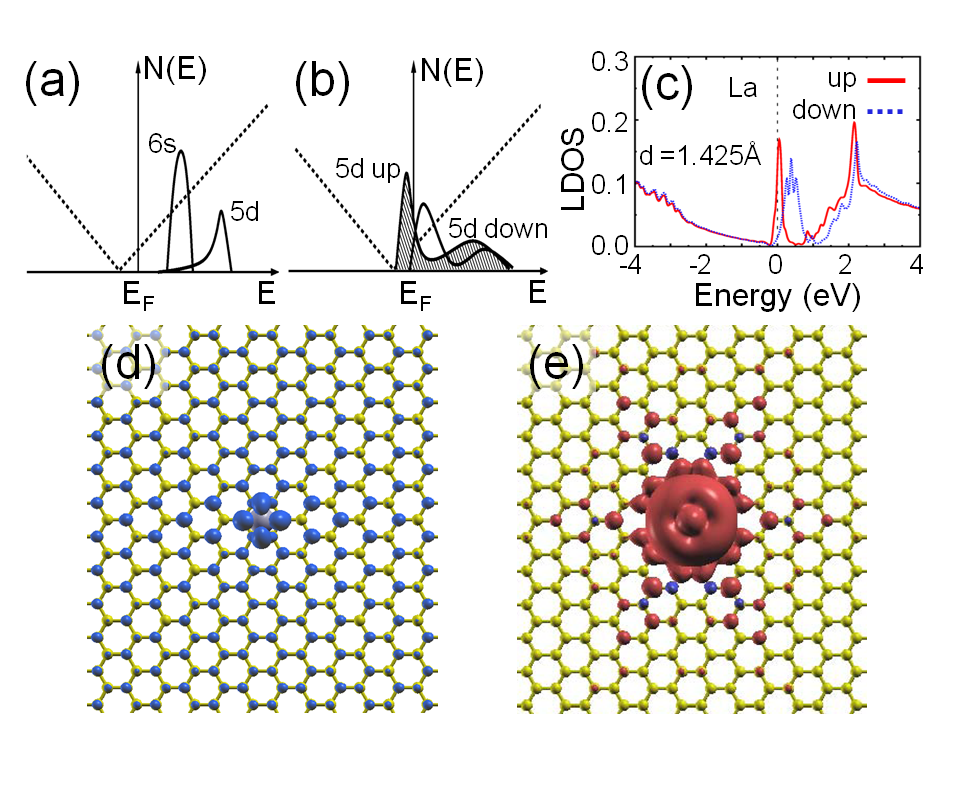}
   \caption{Schematic energy diagrams of DOS of (a) Cs and (b) La on
   graphene. (c) DFT results for the La case at a distance of $d$ = 1.425 \AA. 
   Isosurface plots of (d) charge density associated with a majority spin eigenstate near the Dirac point at $E=-0.18$~eV from 
   the Fermi level (Isovalue = 0.0001 $e$/\AA$^3$) and (e) the total spin density 
   (Isovalues = $\pm0.0001$ $e$/\AA$^3$) in the La-adsorbed graphene.
   In (e), the red and blue colors represent positive and negative values, respectively.
}
   \label{Fig3}
\end{figure}
}

\newcommand{\DOSfig}
{
\begin{figure}[t]
   \centering
   \includegraphics[width=\columnwidth]{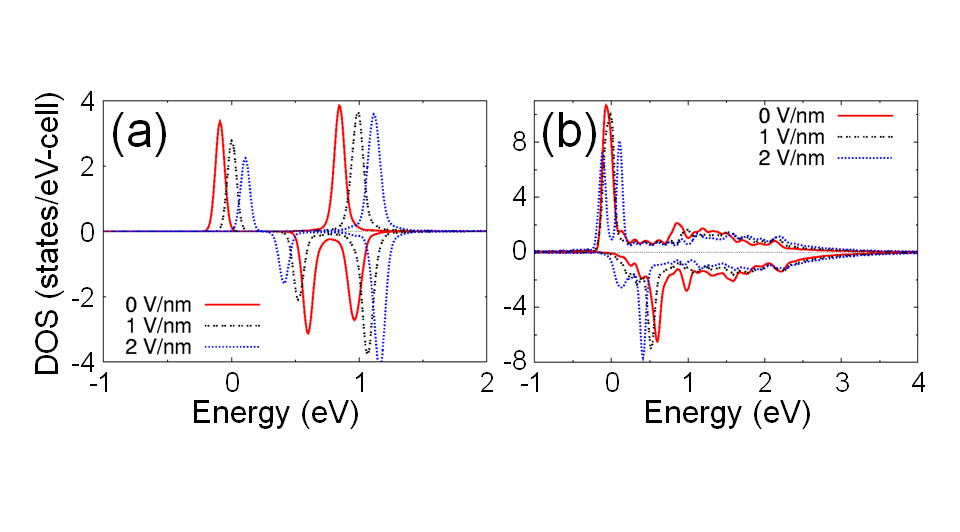}
   \caption{Orbital-resolved DOS only for La (a) La 6$s$ orbital and (b) La 5$d$ 
   orbital in the La-adsorbed graphene with the increased external 
   field.} \label{Fig4}
\end{figure}
}

\begin{document}

\title{Tunable charge donation and spin polarization
of metal adsorbates on graphene using applied electric field}

\author{Jae-Hyeon Parq}
\affiliation{Center for Strongly Correlated Materials Research,
Department of Physics and Astronomy, Seoul National University, Seoul
151-747, Korea}

\author{Jaejun Yu}
\affiliation{Center for Strongly Correlated Materials Research,
Department of Physics and Astronomy, Seoul National University, Seoul
151-747, Korea}

\author{Young-Kyun Kwon}
\affiliation{Department of Physics and Research Institute for Basic
Sciences, Kyung Hee University, Seoul 130-701, Korea}

\author{Gunn Kim}
\email[Electronic mail: ]{gunnkim@khu.ac.kr}
\affiliation{Department of Physics and Research Institute for Basic
Sciences, Kyung Hee University, Seoul 130-701, Korea}

\date{\today}

\begin{abstract}
Metal atoms on graphene, when ionized, can act as a point charge
impurity to probe a charge response of graphene with the Dirac cone
band structure. To understand the microscopic physics of the
metal-atom-induced charge and spin polarization in graphene, we
present scanning tunneling spectroscopy (STS) simulations based on
density functional theory calculations. We find that a Cs atom on
graphene are fully ionized with a significant band bending feature in
the STS, whereas the charge and magnetic states of Ba and La atoms on
graphene appear to be complicated due to orbital hybridization and
Coulomb interaction. By applying an external electric field, we
observe changes in charge donations and spin magnetic moments of the metal adsorbates 
on graphene. 
\end{abstract}


\maketitle

Because of its Dirac cone structure in the energy dispersion,
electrons in graphene behave like two-dimensional (2D) massless Dirac 
fermions at low energies~\cite{1}.
The presence of the Dirac cone structure near the Fermi level plays a
vital role in the determination of physical properties of pristine
graphene. In practical applications, however, the influences of
adatoms and the substrate should also be considered. An adsorbed metal
atom may alter the physical properties, such as the chemical potential
and the effective dielectric constant, of graphene through charge
transfer and orbital hybridization on
graphene~\cite{10,16,17,20}. Recently, a model
based on an assumption of weak coupling between graphene and adatom
was suggested to study 2D quantum-relativistic phenomena on graphene
with a charged metal atom~\cite{21}. This model study predicted that
an atomic collapse-like condensed matter phenomenon may
arise~\cite{21}.

Although several theoretical models have been suggested for charge
impurities in graphene, there has been no attempt to describe a
realistic model with ionized atoms on graphene including the effects
of charge transfer, orbital hybridization, and Coulomb screening
simultaneously. 
Since the amount of charge transfer depends sensitively on
coverage~\cite{24}, which may also can affect adsorption
configurations and bonding characters~\cite{25}, it is necessary to
take a sufficiently large cell for simulations at low coverage.

In this Letter, we present the scanning tunneling spectroscopy (STS)
simulations of graphene with metal adatoms, based on density
functional theory (DFT) calculations. Our STS simulations were 
performed using a large-sized supercell for the low coverage of metal
adsorbates on graphene, and revealed the effects of charge transfer,
orbital hybridization, and Coulomb screening between graphene and
metal adatoms. We selected three metal atoms, cesium (Cs), barium
(Ba), and lanthanum (La), to investigate the dependence on the
electron affinity of adatoms.
We observed band bending near the adatom by monitoring our simulation
data through the van Hove singularity (vHS) lines in the STS plot.
Cs atoms were fully ionized on graphene with a significant band
bending feature in the STS, whereas Ba and La atoms exhibited
complicated electronic and magnetic properties due to orbital
hybridization and Coulomb interactions with graphene.
We also studied the effects of an applied external electric field 
(E-field) on charge donations and magnetic moments of adatoms on 
graphene.

We performed first-principles DFT calculations, using the OpenMX
code~\cite{27}, within the local spin density approximation (LSDA) and
the generalized gradient approximation (GGA)~\cite{28}. General
features of the computational results were almost identical for the two
approximations. All figures presented in this Letter were obtained
from the results of GGA calculations. The OpenMX code employs a linear
combination of pseudo-atomic orbitals (LCPAO)~\cite{29}.
We employed Troullier-Martins-type norm-conserving
pseudopotentials~\cite{30,31} in a factorized separable form with
partial core correction~\cite{32}. Real-space grid
techniques~\cite{33} were used in numerical integration with energy
cutoff up to 200~Ry for LSDA and 300~Ry for GGA.
In our calculations, $8{\times}8{\times}1$ $k$-points were sampled for
calculations with a small unit cell containing 60 carbon atoms, and
$2{\times}2{\times}1$ $k$-points for calculations with a large unit cell
of 800 carbon atoms. The structures were relaxed until the 
Hellman-Feynman force became smaller than $6{\times}10^{-4}$
Hartree/$a_{\rm{B}}$, where $a_{\rm{B}}$ is the Bohr radius.

\strfig

To determine the adsorption site of metal adatoms, we used a
simple orthorhombic unit cell consisting of 60 carbon atoms and one metal
atom, as shown in Fig.~\ref{Fig1}(a).
Among three possible candidate sites (on-top, bridge, and hollow
sites), the hollow site was found to be preferable for all three metal
atoms, which agrees with the previous work~\cite{34} on the Cs atom
where the nature of binding was attributed to ionic
bonding~\cite{10,34,35,36}.
The calcaulated adsorption heights exhibit a trend related to the 
ionic radius of the metal atom~\cite{10}.

\noEfieldtable

Next, we examined charge transfer and magnetic exchange interaction between the adatom
and the graphene layer. The calculated charge and magnetic moment for
each metal atom are listed in Table~\ref{Tab1}, as obtained by the
Voronoi charge analysis~\cite{37}. To ensure the proper description of
the charge response in the graphene layer, we introduced a large
unit cell consisting of approximately 800 carbon atoms and a metal
atom. All the results presented hereafter were
obtained from the large cell calculations.

According to our calculations, a Cs atom donates almost one electron to
graphene to be a Cs$^+$ ion in agreement with an XPS
measurement~\cite{38}. For a Ba or a La atom on graphene, on the other
hand, it turned out that the amount of transferred electrons is 1.2 -- 1.3 $e$. 
It means that such an adsorbed atom is not fully
ionized as observed in previous reports~\cite{10,24}.
Since the electron affinity of graphene is not very high, Ba or La may
not exceed the critical potential strength~\cite{21} on graphene in
the absence of external E-field. Besides, strong orbital
hybridization between graphene states and $d$ electrons of Ba or La
prevents the metal atoms from being fully ionized. This finding
implies that neither Ba nor La acts as a supercritical impurity for an
atomic collapse-like phenomenon~\cite{21}, in contrast to an ealier
study proposing that divalent and trivalent dopants, such as
alkaline-earth and rare-earth metals by used for an experimental
investigation of the phenomenon~\cite{21}.

\STSfig

To simulate the STS data reflecting adatom bonding characters, we
calculated the spatially-resolved local density of states (LDOS).
Figures~\ref{Fig1}(b), \ref{Fig2}(a), and \ref{Fig2}(b) present 2D
simulated STS maps obtained from the calculated LDOS of carbon atoms,
along an armchair line, depicted in Fig.~\ref{Fig1}(a), from the
impurity center to a distant region, the same way used in a previous
study on a carbon nanotube (CNT)~\cite{26}. The Fermi level was
shifted upward by ${\sim}0.2$~eV due to charge donation from the metal
adsorbate to the graphene. 
In the conduction bands, the peak lines corresponding to the vHSs of the Cs-adsorbed graphene form a curve much more prominent 
than those of the Ba- and La-adsorbed cases.
(for clarity, see isovalue lines). On the other hand, the apparent band bending
phenomenon in the valence bands are more clearly shown in Ba- and
La-adsorbed graphene layers than in Cs-doped layer. The differences
between band bending in the Cs-adsorbed case and those in Ba and La
cases originate from the effects of orbital hybridization and the 
Coulomb potential. 
The slope of the asymmetric LDOS around the Dirac point remains intact independent of charge doping, as shown
in Fig.~\ref{Fig1}(c).
Interestingly, the total DOS [Fig.~\ref{Fig1}(d)] is almost
symmetrical around the Dirac point, which implies that distant
graphene states compensate for the asymmetry near the impurity atom.
Similar to a previous prediction by a tight-binding approach~\cite{39},
our calculations show that the Dirac point is robust.
The robustness of the Dirac point is still valid for the cases of Ba
and La, even though there are strong spin polarization and orbital
hybridization effects.

As summarized in Table~\ref{Tab1}, spin magnetic moments on graphene, due to Ba and La atoms, were
0.47 and 1.86 $\mu_{\rm B}$, respectively.
Since the Ba and La atoms are not fully ionized on graphene, the magnetic moments, produced by the
remaining electrons, influence the STS maps, as shown in Fig.~\ref{Fig2}. 
The 6$s$ orbitals of the Ba and La atoms on graphene
are overlapped to the 5$d$ orbitals, and they are partially occupied.
In addition, the relative positions of the up-spin and down-spin
states in the 5$d$ orbitals around the Fermi level bring about the
spin magnetic moments. It is clearly shown that the unoccupied 5$d$
orbitals of Ba and La are strongly hybridized with graphene states in
a wide energy window. The partially occupied atomic states of Ba and
La, especially for up-spin, are located right below the Fermi level,
near the Dirac point. 
Furthermore, the STS maps reveals their covalent bonding characters, which are
encircled in Fig.~\ref{Fig2}. In a previous CNT experiment, such a covalent bonding character between Ba and carbon $p$
orbitals was observed~\cite{40}. For the Cs-adsorbed graphene, in contrast, no
deformation was found in the Dirac point, since Cs is completely
ionized and thus no partially occupied state exists. Moreover,
Fig.~\ref{Fig1}(a), the STS map for the Cs case, does not display the
covalent character spot as expected for the ionic bonding of Cs with C
in graphene.

\Sfig

Ripples of brighter colors are displayed in our STS maps, as indicated
by an arrow in Fig.~\ref{Fig2}(b). Such ripples are the consequence
of strong orbital hybridization between adatoms (Ba and La) and
graphene. 
Most of the ripples are present in the conduction bands; however, the
up-spin valence band for La contains a decaying pattern of $d$-$p$ 
orbital hybridization.
For Ba, on the other hand, a relatively weak decaying pattern appears
in the up-spin valence band, since 5$d$ population in Ba is much smaller
than in La. 
In contrast, the Cs-adsorbed graphene exhibits very weak orbital
hybridization resulting in no ripples as seen in Fig.~\ref{Fig1}(a). 
From charge density analysis, we also found the Friedel
oscillation patterns caused by the charge impurities on graphene (Fig.~\ref{FigS1}).
Our results demonstrate that within $\sim$5 \AA~from the metal adatom, the
amplitudes of the charge density oscillaltions rapidly decrease for
all three metal cases. Besides, asymptotically at large distances
from the adatom, the induced charge density by the impurity charge is
inversely proportional to the cube of the distance from the
impurity~\cite{39,41}.
Our {\em ab initio} calculations, albeit the finite unit cell size, also show similar asymptotic behaviors (Fig.~\ref{FigS2}).
Therefore, it has been confirmed that the effects of Coulomb interaction in metal-atom-adsorbed graphene are {\em insensitive} to orbital hybridization.

\Ediafig

The difference between the behaviors of the Cs atom and those of Ba
and La atoms on graphene is due to the relative positions and widths
of energy levels between the $s$ and $d$ orbitals in the metal atoms.
Figures~\ref{Fig3}(a) and \ref{Fig3}(b) are schematic diagrams of
the LDOSs of Cs and La, respectively, and obviously show this
interesting aspect. Due to charge donation to graphene, the 6s
orbital of Cs becomes completely empty, and is separate from the 5$d$
orbitals with a broad width [Fig.~\ref{Fig3}(a)]. Different from
this feature of Cs, the 6s orbitals of the Ba and La atoms on
graphene are overlapped to the 5$d$ orbitals, and they are partially
occupied. In addition, the relative positions of the up-spin and
down-spin states in the 5$d$ orbitals around the Fermi level bring
about the spin magnetic moments, as seen in Fig.~\ref{Fig3}(b).
Figure~\ref{Fig3}(c) shows DFT results of the local electronic structure in graphene
near the La adsorbate.
In Fig.~\ref{Fig3}(d), a charge density near the Fermi
level shows a combination of 5$d$ orbitals of La with four lobes of adatoms and $p$ orbitals of carbon, 
and corresponds to a rippled pattern in Fig.~\ref{Fig2}(b).
On the other hand, Fig.~\ref{Fig3}(e) reveals that the magnetic moment of the system is mainly from the
atomic spin moment of the metal atom on graphene, and the magnetic
influence of the adatom on graphene seems negligible.

\Efieldtable

Finally, to explore how the ionic charge and magnetic moment of the
metal adatom would change in the presence of external E-field, we
applied external E-field to the La-adsorbed graphene. Using the
gate bias voltage or local E-field from a scanning tunneling
microscope tip, this study can be done in a real measurement. 
As shown in Table~\ref{Tab2}, the La atom moves closer to
the graphene sheet, and donates more electrons. Its ionic charge is,
however, still smaller than +2 $e$ under the applied E-field of up to
2 V/nm. The magnetic moment is reduced due to increasing charge
transfer. The projected DOS of the La atom in Fig.~\ref{Fig4} reveals
the details of electronic state changes. The external field raises the
6$s$ and 5$d$ orbitals for the majority spin, but lowers the 5$d$
orbital for the minority spin. Although the 6$s$ orbital for the
majority spin is occupied in the absence of E-field, it moves up with
the applied field and loses its occupancy. In the 5$d$ orbital,
majority spin occupancy decreases while minority spin occupancy
increases with E-field. In the presence of applied uniform E-field,
the induced charge density by the impurity charge retains the pattern
which is inversely proportional to the cube of the distance from the
impurity. 
If the E-field is in the opposite direction, the spin
magnetic moment of La increases and the amount of donated charge
decreases. Our results indicate that charging and spin magnetic
moments could be tailored by adjusting applied E-fields. Therefore, we
believe that these findings may shed light on the possibility of
electrically tunable (controllable) nano-sized data storage devices.

\DOSfig

In summary, we studied the geometric, electronic, and magnetic
properties of large-area monolayer graphene with a metal adatom using
first-principles calculations. In the two-dimensional STS data, we
found that the differences in the band structure modification of the
Cs and Ba (or La) adatoms originate from the effects of orbital
hybridization as well as electron transfer. Interestingly, the total
DOS is almost symmetrical around the Dirac point, which implies that
far-region graphene states compensate for the asymmetry near the
impurity atom. Although the slopes of the LDOS are asymmetrical around
the Dirac point because of charge doping and orbital hybridization,
the Dirac point is robust. 
By applying uniform electric field, we showed that
field-dependent ionization and changes in spin magnetic moments for
metal adsorbates can occur.

This work was supported by the National Research Foundation
of Korea through the ARP (R17-2008-033-01000-0) (J.Y.) and the Basic Science Research Program through the NRF 
of Korea funded by the Ministry of Education, Science and Technology (No. 2010-0007805) (G.K.).

\end{document}